\documentclass[twocolumn]{webofc}
\usepackage[varg]{txfonts}   
\woctitle{FUSION14}
\begin{document}
\title{Mapping from quasi-elastic scattering to fusion reactions}
%
%

\author{K. Hagino\inst{1,2} \and
        N. Rowley \inst{3}
}

\institute{Department of Physics, Tohoku University, Sendai 980-8578, Japan
\and
Research Center for Electron Photon Science, Tohoku University, 1-2-1 Mikamine, Sendai 982-0826, Japan
\and
Institut de Physique Nucl\'{e}aire, UMR 8608, CNRS-IN2P3 et Universit\'{e}
de Paris Sud, 91406 Orsay Cedex, France
          }

\abstract{%
The fusion barrier distribution has provided a nice representation for the 
channel coupling effects on heavy-ion fusion reactions at energies 
around the Coulomb barrier. 
Here we discuss how one can extract the same representation using 
the so called sum-of-differences (SOD) method with quasi-elastic 
scattering cross sections. 
In contrast to the conventional quasi-elastic barrier 
distribution, the SOD barrier distribution has an advantage in that 
it can be applied both to non-symmetric and symmetric systems. 
It is also the case that 
the correspondence to the fusion barrier distribution is much 
better than the quasi-elastic barrier distribution. We demonstrate its 
usefulness by studying $^{16}$O+$^{144}$Sm, $^{58}$Ni+$^{58}$Ni, and 
$^{12}$C+$^{12}$C systems. 
}
\maketitle
\section{Introduction}

The fusion barrier 
distribution \cite{RSS91} has by now been a standard representation for 
heavy-ion fusion cross sections at energies around the Coulomb 
barrier \cite{DHRS98,LDH95,BT98,HT12}. 
It can directly be obtained from measured fusion cross sections 
$\sigma_{\rm fus}$ as 
$D_{\rm fus}(E)=d^2(E\sigma_{\rm fus})/dE^2$, that is, 
by taking the second derivative of 
$E\sigma_{\rm fus}$ with respect to $E$, 
where $E$ is the incident energy in the center 
of mass frame \cite{RSS91}. 
The fusion barrier distribution has been 
experimentally extracted for a number of systems. 
These experimental data have indicated that the fusion barrier 
distribution is sensitive to the channel coupling effects and it thus 
provides a powerful method to study the energy dependence of fusion 
cross sections at subbarrier energies \cite{DHRS98,LDH95}. 

A similar barrier distribution has been proposed also for quasi-elastic 
scattering cross sections \cite{TDH94}, 
that is, a sum of elastic, inelastic, and transfer cross sections. 
The quasi-elastic barrier distribution is defined as 
the first derivative of the ratio of the quasi-elastic cross section 
$\sigma_{\rm qel}$ to the Rutherford cross section $\sigma_{\rm Ruth}$, that is, 
$D_{\rm qel}(E)=-d(d\sigma_{\rm qel}/d\sigma_{\rm Ruth})/dE$, in which 
both $\sigma_{\rm qel}$ and $\sigma_{\rm Ruth}$ are evaluated at 
the scattering 
angle of $\theta_{\rm c.m.}=\pi$. 
It has been demonstrated both experimentally and theoretically that 
the quasielastic barrier distribution behaves similarly 
to the fusion barrier distribution \cite{TDH94,HR04}. 

\begin{figure}
\centering
\includegraphics[scale=0.45,clip]{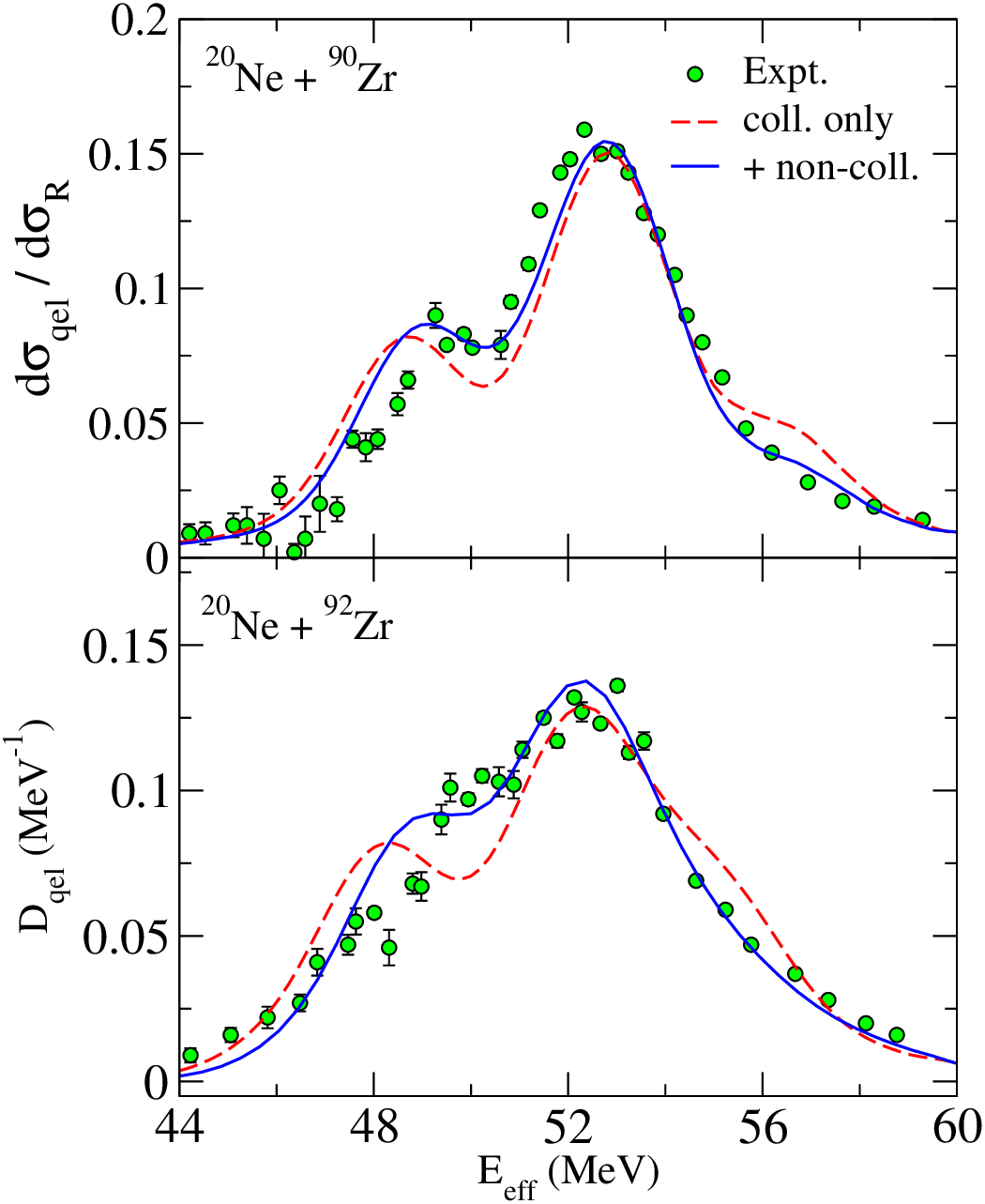}
\caption{The quasi-elastic barrier distributions for the 
$^{20}$Ne+$^{90}$Zr (the upper panel) and the 
$^{20}$Ne+$^{92}$Zr (the lower panel) systems. 
These are evaluated at the scattering angle of 
$\theta_{\rm lab}$ = 150$^\circ$ and are 
plotted as a function of effective energy 
defined by 
$E_{\rm eff}= 2E\sin(\theta_{\rm c.m.}/2)/(1+\sin(\theta_{\rm c.m.}/2))$.
The dashed lines show the results of the coupled-channels calculations 
with the collective excitations in the projectile and the target nuclei, 
while the solid lines take in addition the non-collective excitations in the 
target nuclei into account with a random matrix model. 
The experimental data are taken from Ref. \cite{Piasecki}. 
}
\end{figure}

As an example of recent applications of the quasi-elastic 
barrier distribution, Fig. 1 shows a comparison between the 
quasi-elastic barrier distribution for the 
$^{20}$Ne + $^{90}$Zr system and that for 
the $^{20}$Ne + $^{92}$Zr systems \cite{YHR13}. 
One striking thing is that the experimental 
quasi-elastic barrier distribution 
for the $^{20}$Ne + $^{92}$Zr system is much more smeared 
than that 
for the $^{20}$Ne + $^{90}$Zr system \cite{Piasecki}. 
The dashed lines in the figure 
show the results of the coupled-channels calculations 
that include the rotational excitations in $^{20}$Ne as well 
as the collective phonon excitations in $^{90,92}$Zr. 
This calculation reproduces the experimental data for the 
$^{20}$Ne + $^{90}$Zr system but not for the 
$^{20}$Ne + $^{92}$Zr system.  
The solid lines, on the other hand, take into account 
also the non-collective excitations in $^{90,92}$Zr with a random matrix model. 
One can see that the smearing of quasi-elastic barrier distribution 
for the $^{20}$Ne + $^{92}$Zr system is now well reproduced by 
the non-collective excitations of the $^{92}$Zr nucleus, whose level 
density is much larger than that of $^{90}$Zr due to the two extra neutrons 
outside the $N=50$ shell closure. 

\section{The sum-of-differences method}

While a similar barrier distribution can be obtained both from fusion and 
from quasi-elastic cross sections, 
the quasi-elastic barrier distribution has several experimental 
advantages over the fusion barrier distribution, such as the fact 
that less accuracy is required in the data for taking the first 
derivative rather than the second derivative \cite{HR04}. 
However, at the same time, it also has two drawbacks. Firstly, the 
quasi-elastic barrier distribution 
is somewhat smeared and thus less sensitive to
the nuclear structure effect compared to the fusion barrier distribution. 
This is due to the effect of nuclear distortion of 
a classical trajectory \cite{HR04}, 
which is not taken into account in the definition 
of quasi-elastic barrier distribution. This sometimes leads to a large 
difference between the fusion and the quasi-elastic barrier distributions, 
a well known example being the $^{16}$O+$^{144}$Sm system \cite{TDH94,ZH08}. 
Secondly, the quasi-elastic barrier distribution cannot be applied to 
symmetric systems, because the quasi-elastic cross sections diverge at 
$\theta_{\rm c.m.}=\pi$ due to the (anti-)symmetrization effect. 

In order to avoid these drawbacks, we here propose to use the so called 
sum-of-differences (SOD) method. This method was proposed many years ago by 
Holdeman and Thaler \cite{HT65}, who argued that the reaction cross section 
$\sigma_{\rm R}$ is obtained from the elastic cross section $\sigma_{\rm el}$ as,
\begin{equation}
\sigma_{\rm R} \sim 2\pi\int^\pi_{\theta_{\rm min}}
\left(\frac{d\sigma_{\rm Ruth}}{d\Omega} - \frac{d\sigma_{\rm el}}{d\Omega}\right)
\sin\theta d\theta,
\label{SOD0}
\end{equation}
where $\theta_{\rm min}$ is an angle such that the difference between 
the elastic and the Rutherford cross sections is negligibly small for 
$\theta < \theta_{\rm min}$. 
See also Refs. \cite{M83,Hussein,BA85,L87,WMD77,OHHS79,Yamaya}. 
Since the reaction cross section $\sigma_{\rm R}$ is given by 
$\sigma_{\rm R}=\sigma_{\rm fus}+\sigma_{\rm inel}+\sigma_{\rm tr}$, where 
$\sigma_{\rm inel}$ and $\sigma_{\rm tr}$ are inelastic and transfer 
cross sections, respectively, Eq. (\ref{SOD0}) can be transformed 
also to \cite{HR14} 
\begin{equation}
\sigma_{\rm fus} \sim 2\pi\int^\pi_{\theta_{\rm min}}
\left(\frac{d\sigma_{\rm Ruth}}{d\Omega} - \frac{d\sigma_{\rm qel}}{d\Omega}\right)
\sin\theta d\theta,
\label{SOD}
\end{equation}
where the quasi-elastic cross section is given by 
$\sigma_{\rm qel} = \sigma_{\rm el} +\sigma_{\rm inel}+\sigma_{\rm tr}$. 
We define the SOD barrier distribution, $D_{\rm SOD}$, as the fusion barrier 
distribution with fusion cross sections so obtained. 

For symmetric systems with spin-zero bosons, Eq. (\ref{SOD}) is modified 
as \cite{HR14,OTV91} 
\begin{equation}
\sigma_{\rm fus} \sim 2\pi\int^{\theta_{\rm max}}_{\pi/2}
\left(\frac{d\sigma_{\rm Mott}}{d\Omega} - \frac{d\sigma_{\rm qel}}{d\Omega}\right)
\sin\theta d\theta,
\label{SOD-sym}
\end{equation}
where $\sigma_{\rm Mott}$ is the Mott cross section. 

\section{SOD barrier distributions}

\subsection{Non-symmetric systems}

Let us now examine the performance of the SOD barrier distribution. 
We first consider the $^{16}$O+$^{144}$Sm system as an example for 
non-symmetric systems. We take into account the one phonon couplings 
with the quadrupole and octupole phonons in $^{144}$Sm while $^{16}$O is 
assumed to be inert. The coupled-channels equations are solved with a 
version of the computer code {\tt CCFULL} \cite{HRK99}. We use an imaginary 
potential with the Woods-Saxon form 
for the internuclear potential, whose imaginary part is well confined 
inside the Coulomb barrier. 

\begin{figure}
\centering
\includegraphics[scale=0.45,clip]{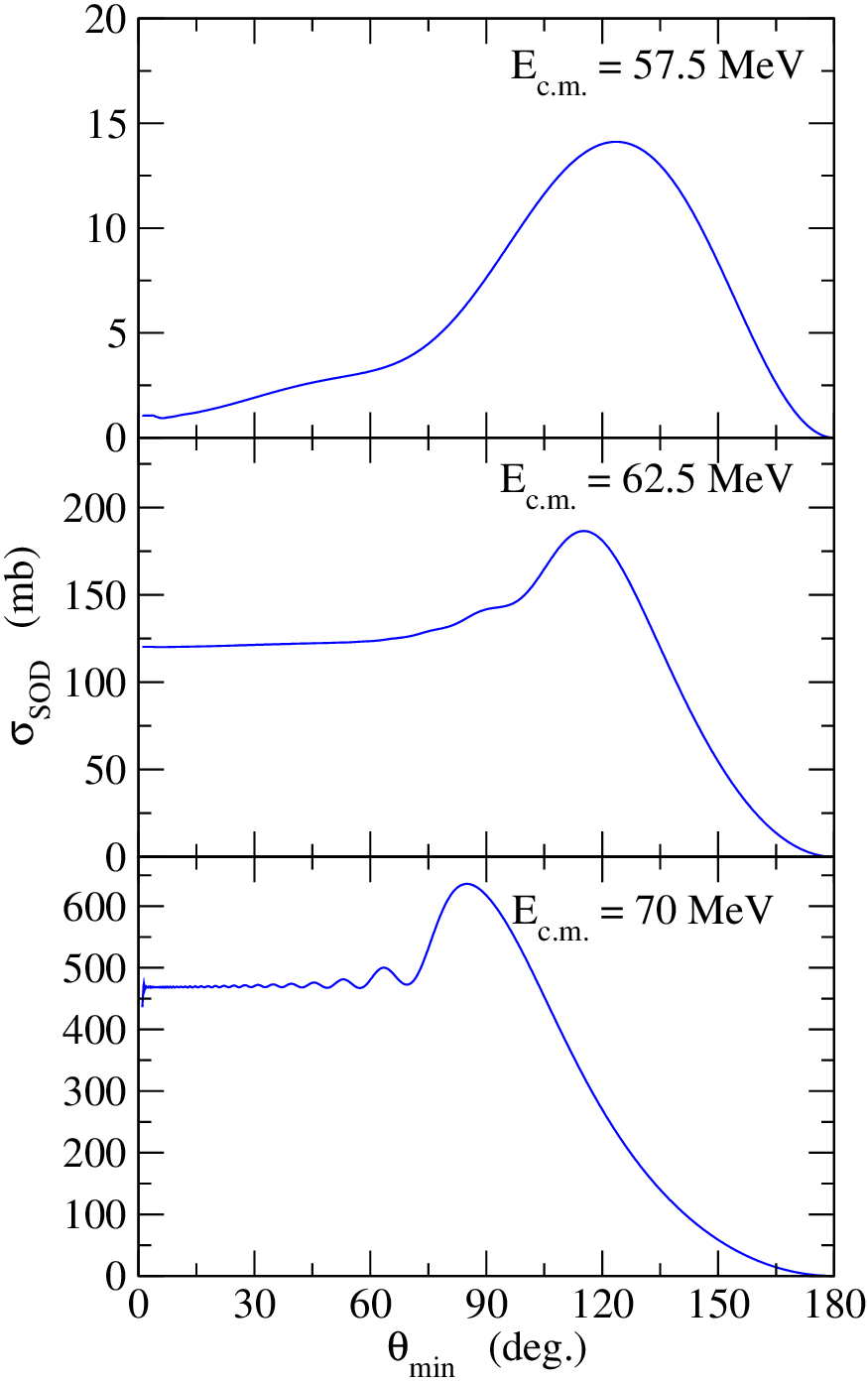}
\caption{The fusion cross sections obtained with Eq. (\ref{SOD}) 
for the $^{16}$O+$^{144}$Sm system as a function of the minimum 
angle $\theta_{\min}$. The one phonon excitations 
of the quadrupole and octupole modes in $^{144}$Sm are taken into 
account. 
}
\end{figure}

Figure 2 shows the fusion cross sections obtained with Eq. (\ref{SOD}) 
at three different incident energies 
as a function of the minimum angle, $\theta_{\rm min}$. 
One can see that the fusion cross section is insensitive to the choice 
of $\theta_{\rm min}$, as long as it is smaller than about 40 degrees. 
The dots in the upper panel of Fig. 3 shows the fusion excitation function 
evaluated with $\theta_{\rm min}=20^\circ$. 
For a comparison, 
the solid line shows the exact fusion cross sections 
directly obtained with the coupled-channels 
calculations. 
The fusion cross sections obtained with the SOD method are almost 
indistinguishable from the exact fusion cross sections in the scale 
shown in the figure. 

The lower panel of Fig. 3 shows the fusion 
barrier distributions obtained by taking the second derivative of 
the fusion cross sections shown in the upper panel. 
We normalize the exact fusion barrier distribution 
between 55 $\leq E_{\rm c.m.} \leq$ 70 MeV and multiply the same normalization 
factor to the SOD barrier distribution. 
One can see that the SOD 
barrier distribution well 
reproduces the exact fusion barrier distribution, except for 
the high energy region. 
The deviation in this energy region is due to the finite value of 
$\theta_{\rm min}$, as one can infer from the small oscillations shown in the 
bottom panel of Fig. 2. In fact, if one takes a larger value of 
$\theta_{\rm min}$, {\it e.g.,} 
$\theta_{\rm min}=30^\circ$, the deviation becomes even 
larger although the main 
structure of fusion barrier distribution is still reproduced. 

For a comparison, 
Fig. 3 also shows the corresponding quasi-elastic barrier distribution by the 
dotted line. 
While the SOD barrier distribution reproduces the exact fusion 
barrier distribution almost perfectly, the quasi-elastic barrier distribution 
does not coincide with the fusion barrier distribution, although the main 
structure is qualitatively reproduced. The main peak is somewhat lowered 
and the 
peak position is slightly shifted towards a lower energy. At the same time, 
the peaks have a larger tail on the low energy side. 
These features can be attributed to the nuclear distortion effect, as has been 
demonstrated in Fig. 1 in Ref. \cite{HR04}. 
Evidently, the SOD barrier distribution has a much better correspondence 
to the fusion barrier distribution as compared with the quasi-elastic 
barrier distribution. 

Of course, at energies well below the Coulomb 
barrier, the SOD cross sections are difficult to obtain, both 
theoretically and experimentally, since the 
quasi-elastic and the Rutherford cross sections are almost the same. 
Fortunately, this energy region does not contribute to the structure of 
fusion barrier distribution, as shown in the lower panel of Fig. 3. 

\begin{figure}
\centering
\includegraphics[scale=0.45,clip]{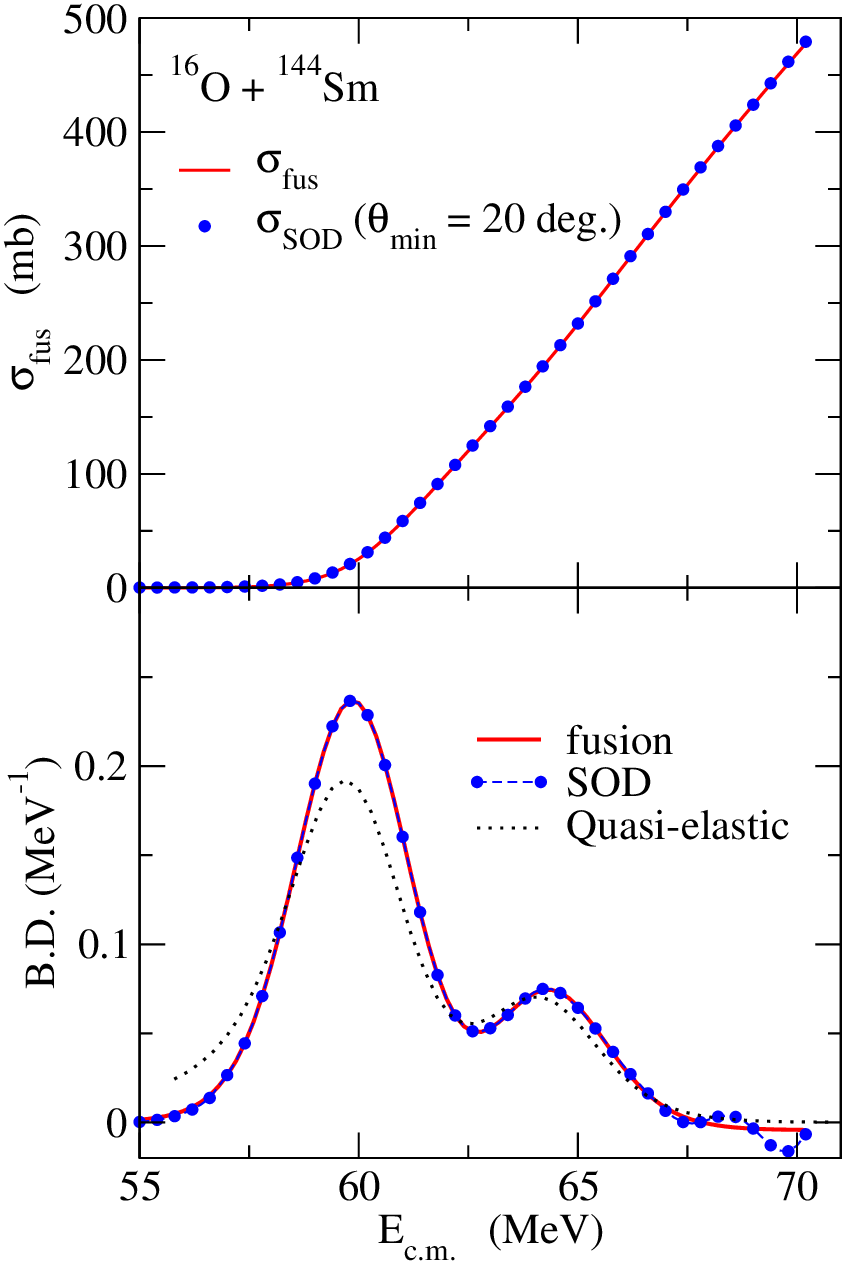}
\caption{(the upper panel) The fusion excitation function for the 
$^{16}$O+$^{144}$Sm system. The solid line is obtained with the 
coupled-channels calculations with the one phonon quadrupole and 
octupole excitations in $^{144}$Sm. The dots are obtained with 
Eq. (\ref{SOD}) with $\theta_{\rm min}=20^\circ$ 
using the corresponding quasi-elastic cross sections. 
(the lower panel) The corresponding fusion barrier distributions 
normalized in the energy range of 
55 $\leq E_{\rm c.m.} \leq$ 70 MeV. 
The quasi-elastic barrier distribution is also shown by the dotted line.}
\end{figure}

\subsection{Symmetric systems}

Let us next discuss symmetric systems. 
For such systems, 
the elastic cross section is given by
\begin{equation}
\frac{d\sigma_{\rm el}}{d\Omega}=|f(\theta)\pm f(\pi-\theta)|^2,
\end{equation}
where $f(\theta)$ is the scattering amplitude for elastic scattering 
and the sign on the right hand side of this equation 
depends on whether the spatial part 
of wave function is symmetric (+) or anti-symmetric ($-$) 
with respect to the exchange 
between the projectile and the target. 
Apparently this cross section diverges at $\theta=\pi$, as the Rutherford 
cross section diverges at $\theta=0$. 
For symmetric systems, 
this prevents the 
use of the quasi-elastic barrier distribution, which is 
defined at $\theta\sim\pi$. 
Even in that situation, however, 
one can still map quasi-elastic cross sections 
to fusion cross sections by using the SOD method given by 
Eq. (\ref{SOD-sym}). 

\begin{figure}
\centering
\includegraphics[scale=0.45,clip]{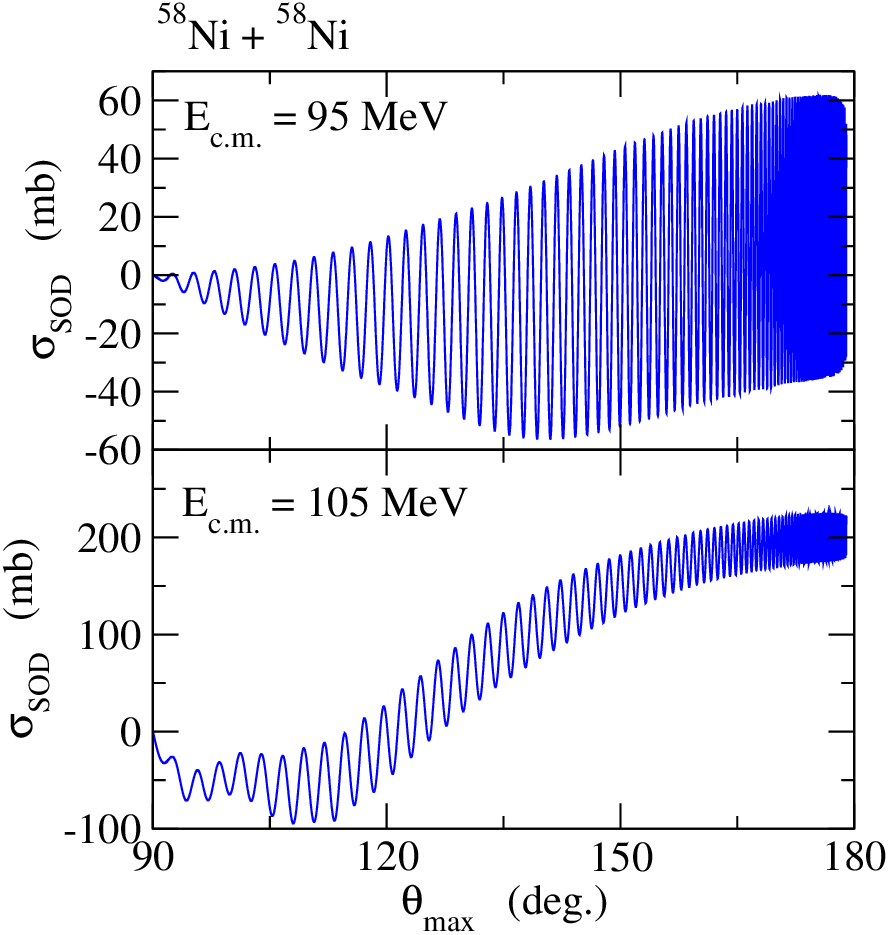}
\caption{The fusion cross sections obtained with Eq. (\ref{SOD-sym}) 
for the $^{58}$Ni+$^{58}$Ni system as a function of the maximum 
angle $\theta_{\max}$. The one quadrupole phonon excitations 
in the projectile and the target nuclei, as well as the mutual 
excitation channel, are taken into account. 
}
\end{figure}

In order to demonstrate the performance of the SOD barrier distribution 
for symmetric systems, we consider the $^{58}$Ni + $^{58}$Ni system. 
We include the one quadrupole phonon excitations at 1.45 MeV 
in the projectile and the 
target nuclei as well as the mutual excitation channel. 
Fig. 4 shows the fusion cross sections obtained with Eq. (\ref{SOD-sym}) 
as a function of $\theta_{\rm max}$ at two different incident energies. 
In contrast to the non-symmetric system shown in Fig. 2, the SOD cross 
sections oscillates rapidly as a function of $\theta_{\rm max}$. 
This of course is due to the oscillatory nature of the Mott 
scattering \cite{HAT07}, 
which is caused by the interference between the forward and the backward 
amplitudes, that is, $f(\theta)$ and $f(\pi-\theta)$, respectively. 

Since it is meaningless to evaluate the SOD cross sections at a fixed 
value of $\theta_{\rm max}$, due partly to a finite angle resolution in 
actual experiments,  we obtain fusion cross sections for this system by 
averaging the SOD cross sections within a small range of $\theta_{\rm max}$. 
The dots in the upper panel of Fig. 5 are so obtained by averaging 
an upper integration limit $\theta_{\rm max}$ between 
176.5$^\circ$ and 179.5$^\circ$. 
Again, these reproduce almost perfectly the 
exact fusion cross sections shown by the solid 
line. The lower panel of Fig. 5 shows the corresponding fusion barrier 
distributions. Although the agreement 
is slightly worse than that for the non-symmetric 
system shown in Fig. 3, one can still see that 
the SOD barrier distribution well reproduces the 
exact fusion barrier distribution. 

\begin{figure}
\centering
\includegraphics[scale=0.45,clip]{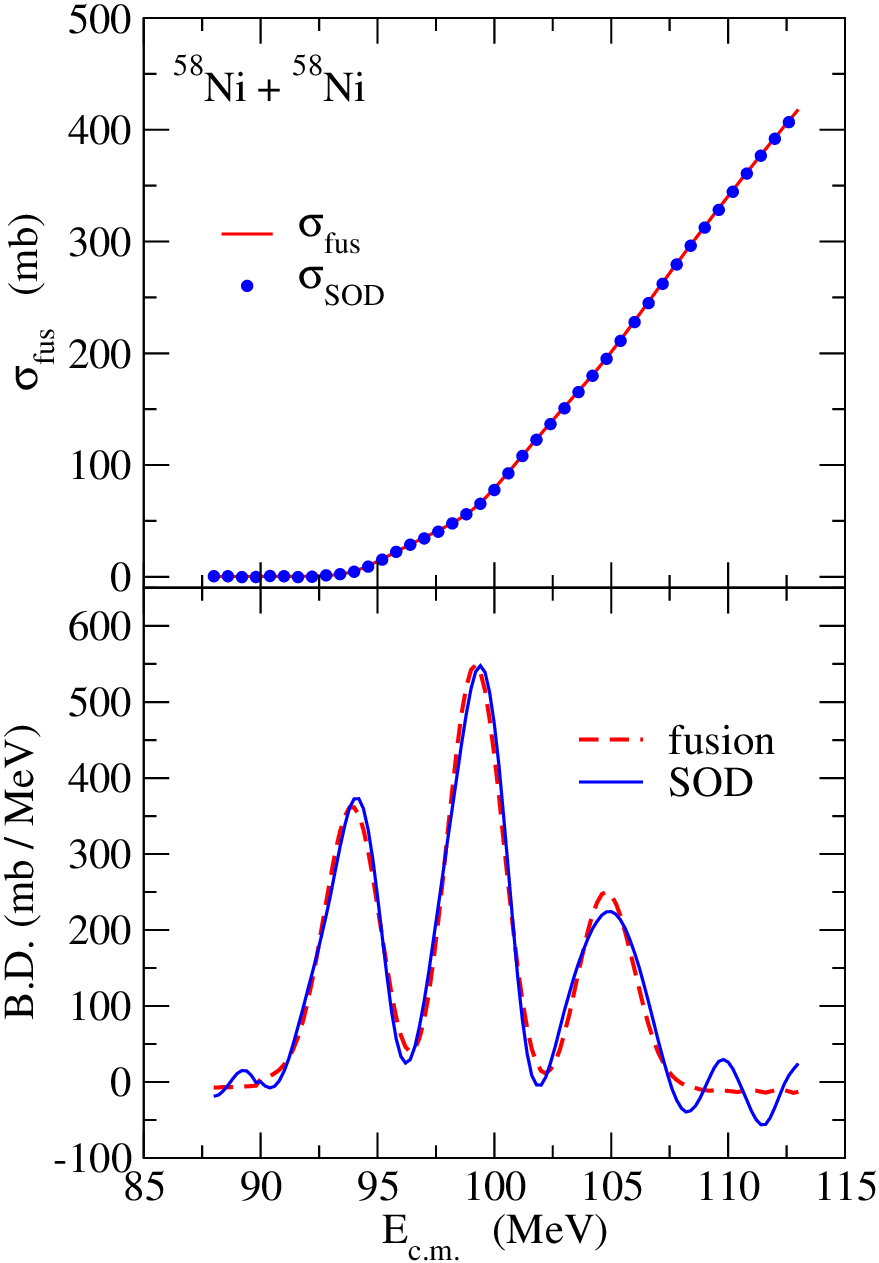}
\caption{(the upper panel) The fusion excitation function for the 
$^{58}$Ni+$^{58}$Ni system. The solid line is obtained with the 
coupled-channels calculations with the one quadrupole phonon 
excitations in the projectile and the target nuclei, as well as the 
mutual excitation channel. 
The dots show the results based on 
Eq. (\ref{SOD-sym}) 
with the corresponding quasi-elastic cross sections. 
These are obtained by averaging 
over an upper integration limit of 
the SOD cross sections between 
$\theta_{\rm max} =176.5^\circ$ and 179.5$^\circ$. 
(the lower panel) The corresponding fusion barrier distributions. } 
\end{figure}

\section{Fusion oscillations in light symmetric systems}

An interesting application of the SOD method may be to light symmetric 
systems, such as $^{12}$C+$^{12}$C and $^{16}$O+$^{16}$O. 
Fig. 6 shows the experimental fusion cross sections for the 
$^{12}$C+$^{12}$C system. See Fig. 1 in Ref. \cite{SKUO13} 
for a similar figure for the $^{16}$O+$^{16}$O system. 
One can see a large scatter in the experimental data. This is partly due to 
systematic errors in the experiments 
and partly due to missing evaporation channels. 
That is, some of these experimental data were obtained by measuring 
$\gamma$-rays from evaporation residues, where it may be hard to detect 
all the evaporation channels. 
The SOD method provides a good alternative way to experimentally 
determine fusion cross sections, as 
a detection of quasi-elastic 
cross sections is presumably 
much simpler than a direct detection of evaporation 
residues, although measurements at many scattering angles may be time 
consuming. 

\begin{figure}
\centering
\includegraphics[scale=0.45,clip]{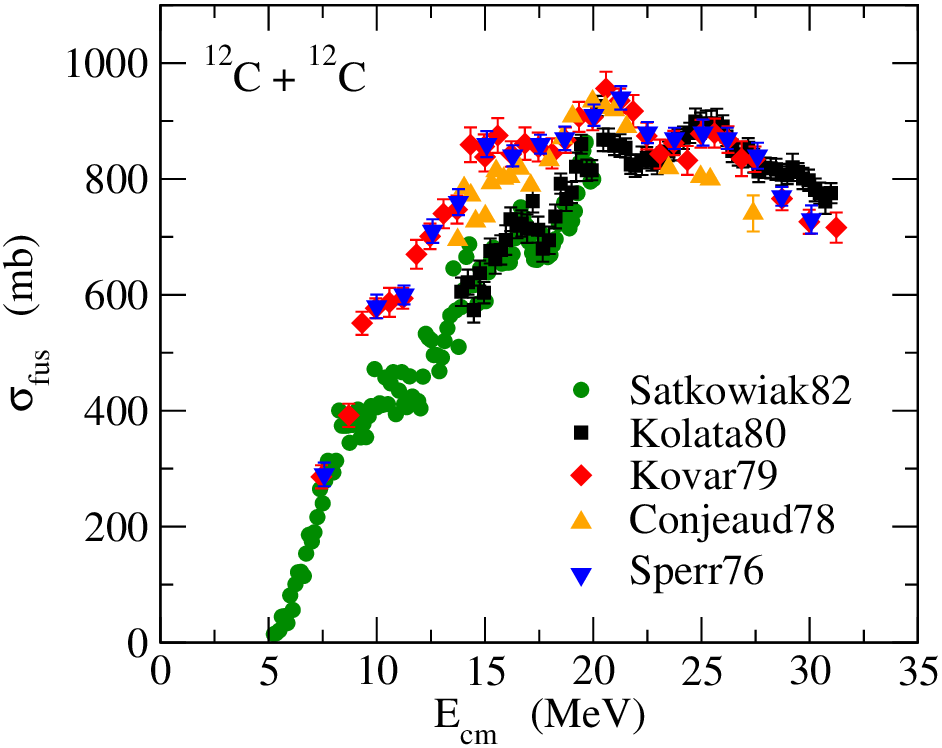}
\caption{The experimental fusion cross sections for the $^{12}$C+$^{12}$C 
system, taken from Refs.\cite{SDKX82,Kolata80,Kovar79,Conjeaud78,Sperr76}. }
\end{figure}

\begin{figure}
\centering
\includegraphics[scale=0.45,clip]{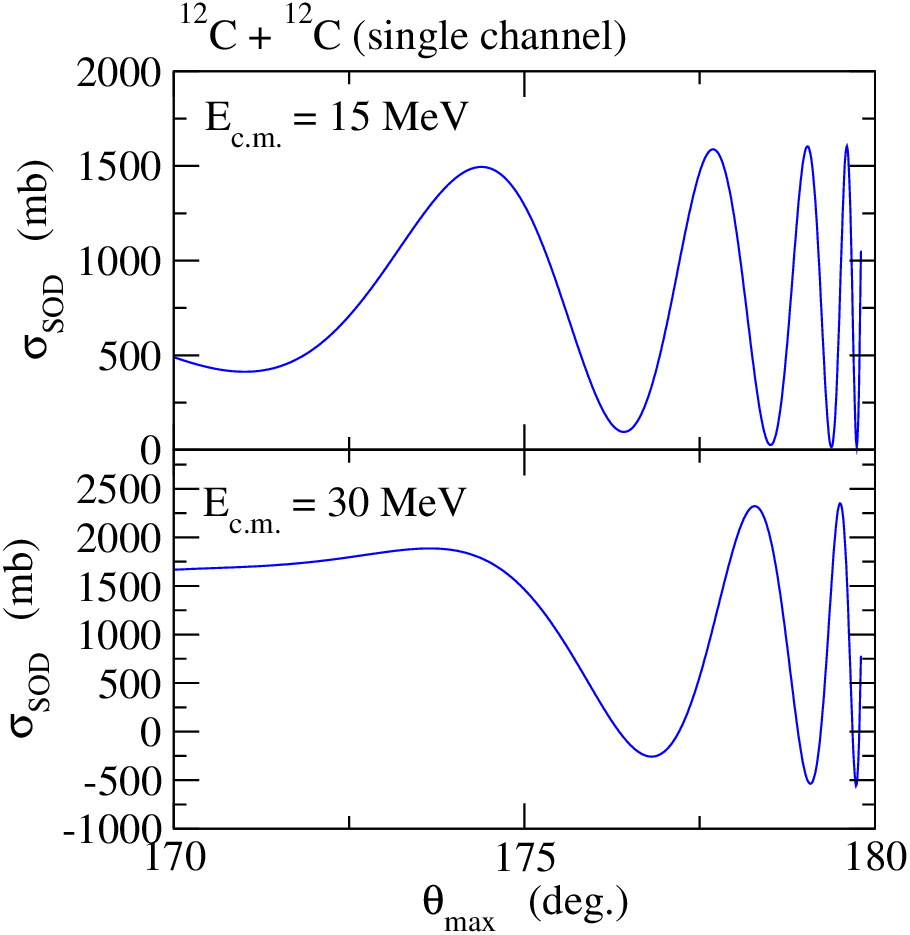}
\caption{The same as Fig. 4, but for the 
$^{12}$C+$^{12}$C system with single-channel calculations.}
\end{figure}

\begin{figure}
\centering
\includegraphics[scale=0.45,clip]{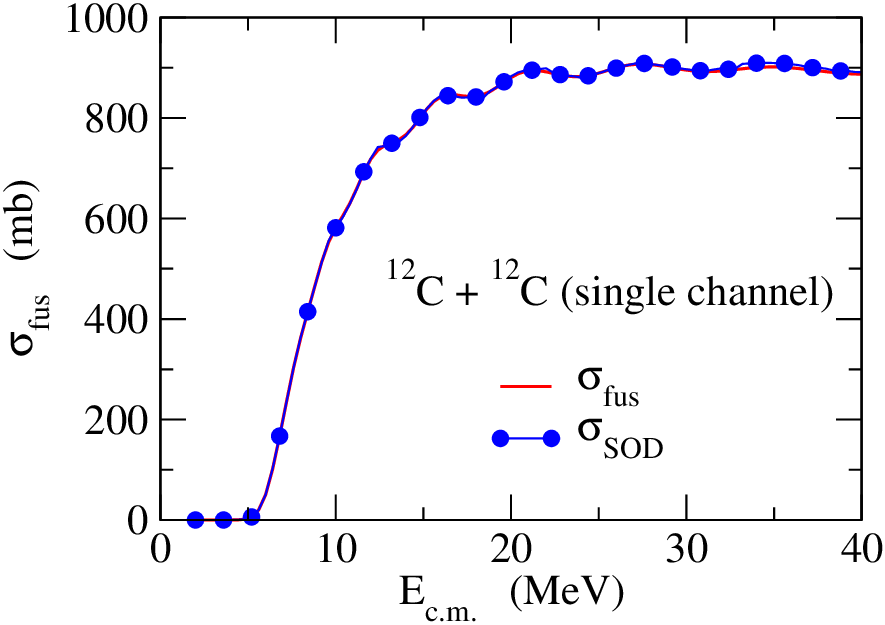}
\caption{The same as the upper panel of Fig. 5, but for the 
$^{12}$C+$^{12}$C system with single-channel calculations.}
\end{figure}

Figure 7 shows the SOD cross sections for the $^{12}$C+$^{12}$C system 
obtained with Eq. (\ref{SOD-sym}) as a function of $\theta_{\rm max}$. 
To this end, we carry out single-channel calculations with an exponential 
potential with the diffuseness parameter of $a$=0.8 fm and the depth 
parameter of $V_0=-8028.5$ MeV. 
Since the Sommerfeld parameter is smaller, the SOD cross sections are much 
less oscillatory for this system as compared to the 
$^{58}$Ni+$^{58}$Ni system shown in Fig. 4. 
We then obtain the fusion cross sections by averaging the maximum and minimum 
in the SOD cross section close to $\theta_{\rm max}=\pi$. 
The fusion cross sections so obtained are shown in Fig. 8 by the dots. 
As for the $^{58}$Ni+$^{58}$Ni system shown in Fig. 5, these calculations well 
reproduce the exact fusion cross sections, including the oscillatory character. 

The fusion oscillations observed in light symmetric systems can be 
interpreted as due to the addition of successive individual partial 
waves as the energy increases \cite{PRL83,KKR83} (see also 
Refs.\cite{SKUO13,E12,W12}). This effect is 
enhanced in systems with identical spin-zero bosons. In such systems, 
only even-partial waves contribute, increasing the energy spacing 
between successive contributing angular momenta. 
Within the parabolic approximation, 
one can derive 
the oscillatory part of fusion cross sections as \cite{PRL83,HT12}
\begin{equation}
\sigma_{\rm osc}(E)=\pm 2\pi R_b^2\frac{\hbar\Omega}{E}\exp\left(-
\frac{\pi\mu R_b^2\hbar\Omega}{(2l_g+1)\hbar^2}\right)\sin(\pi l_g), 
\label{osc}
\end{equation}
where the positive (negative) 
sign corresponds to the spatially symmetric (asymmetric) case with respect 
to the exchange of the projectile and the target nuclei. 
In this equation, $R_b$ and $\hbar\Omega$ are the barrier position and the 
curvature of the Coulomb barrier, respectively, $\mu$ is the reduced 
mass of the system, and $l_g$ is the grazing angular momentum. 
The oscillatory part of fusion cross sections, $\sigma_{\rm osc}$, is 
to be added to the smooth part of fusion cross sections given by the 
well-known Wong formula \cite{Wong}. 
Notice that, although the original form of the 
Wong formula does not work for light systems, 
because the angular momentum dependence of $R_b$ and $\hbar\Omega$ is 
strong, the Wong formula itself as well as 
Eq. (\ref{osc}) still work as long as 
the barrier parameters 
$R_b$ and $\hbar\Omega$ are evaluated at the grazing angle $l_g$ rather 
than at $l=0$ \cite{RKL89,RH14}. 

\begin{figure}
\centering
\includegraphics[scale=0.45,clip]{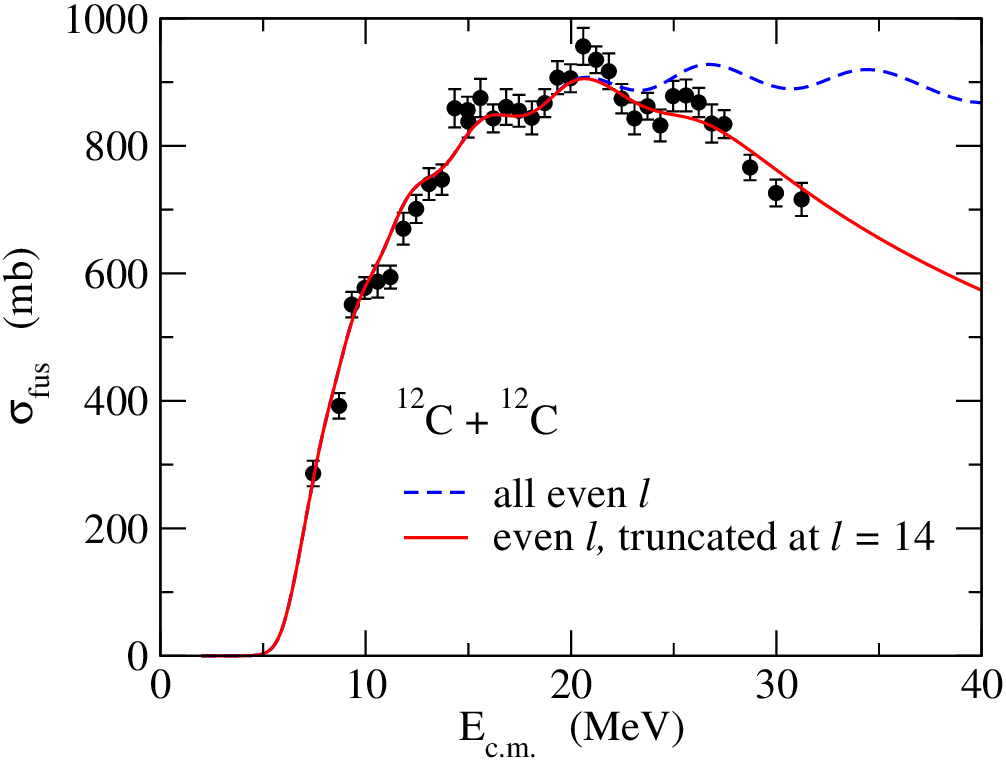}
\caption{
Fusion excitation functions for the $^{12}$C+$^{12}$C system 
obtained with the single-channel calculations. 
An exponential potential with the diffuseness parameter of $a=0.8$ fm 
is employed for the internuclear potential. 
The dashed line is obtained by including all 
even partial waves, while the solid 
line is obtained by truncating the even partial waves at $l=14$. 
The experimental data are taken from 
Refs.\cite{Kovar79,Sperr76}. }
\end{figure}

Lastly, we present our potential model fit to the experimental cross 
sections for the $^{12}$C+$^{12}$C system. 
The dashed line in Fig. 9 shows the fusion cross sections obtained by 
including all even partial waves. It reproduces fairly well the experimental 
data up to around $E_{\rm c.m.}$=25 MeV. 
At higher energies, however, the calculation 
overestimates fusion cross sections, a possible explanation for this being 
the failure of higher partial waves to fuse \cite{RH14}. 
As shown by the solid line in Fig. 9, 
one can in fact obtain a nice fit to the data 
by phenomenologically 
reducing the penetrability of the Coulomb barrier for $l=14$ 
by a factor of 2 and setting the penetrabilities for all the higher partial 
waves to be zero. 
Such an assumption may be justified given that 
the excitation energy for the compound nucleus, $^{24}$Mg, at 
energies corresponding to the barrier height for each partial wave 
is estimated to be below $E_{\rm yrast}+S_n$ and $E_{\rm yrast}+S_p$ for 
high partial waves, where 
$S_n$ and $S_p$ are one neutron and one proton 
separation energies, respectively, 
and $E_{\rm yrast}$ is the yrast energy. 
Particle emission decays from the 
compound nucleus formed at such low non-collective 
excitation energies 
do not take place and the compound nucleus decays only by fission 
(that is, a reseparation into the entrance channel) or by the 
relatively slow process of $\gamma$ decay, hindering the measured 
fusion cross sections. 

\section{Summary}

We have advocated a use of the sum-of-differences (SOD) method as 
a promising method to deduce fusion cross sections from quasi-elastic 
cross sections. 
The barrier distribution with the fusion cross sections 
obtained by the SOD method, 
that is, the SOD barrier distribution, 
has a much better correspondence to the fusion barrier distribution 
as compared to the conventional quasi-elastic barrier distribution. Moreover, 
the SOD barrier distribution can be applied also to symmetric systems, whereas 
the quasi-elastic barrier distribution is not applicable due to the divergence 
of quasi-elastic cross sections at the scattering angle of $\pi$. 
We have demonstrated these attractive 
features by carrying out the coupled-channels 
calculations for the $^{16}$O+$^{144}$Sm, $^{58}$Ni+$^{58}$Ni, 
and $^{12}$C+$^{12}$C systems. 
The price to pay is that quasi-elastic cross sections have to be measured 
over many scattering angles with a small angle step so that one can 
perform the angle integral, whereas the 
conventional quasi-elastic barrier distribution requires experimental data 
at only a few backward angles. An interesting application of the SOD method 
may be to light symmetric systems, where fusion oscillations have been 
observed. For such systems, the experimental fusion cross sections 
often show significant discrepancies among each other, due partly to 
a difficulty 
to detect all the evaporation channels. 
For such systems, compared to a direct detection of evaporation residues, 
it might be easier to measure 
quasi-elastic cross sections, with which one can construct fusion 
cross sections using the SOD method. 

\section*{Acknowledgment}

We thank M. Dasgupta and D.J. Hinde for useful discussions. 
This work was supported by
JSPS KAKENHI Grand Number 25105503.

\end{document}